**Title:** *"Detecting the Onset and Progression of Spinodal Decomposition using Transient Grating Spectroscopy"*

**Authors & affiliations:**

Maxwell Rae[a], Merrill K. Chiang[a], Mahmudul Islam[a], Angus P. C. Wylie[b], Avery Nguyen[b], Myles Stapelberg[b], Saleem A. Al Dajani[c], Kristýna Repček[d], Tomáš Grabec[d], Abigail Kaplan[b], Austin Wang[e], Austin J. Akey[e], Rodrigo Freitas[a], Michael P. Short[b]

[a]Dept. of Materials Science and Engineering, Massachusetts Institute of Technology, Cambridge, MA 02139, USA
[b]Dept. of Nuclear Science and Engineering, Massachusetts Institute of Technology, Cambridge, MA 02139, USA
[c]Department of Civil and Environmental Engineering, Massachusetts Institute of Technology, 77 Massachusetts Avenue, Cambridge, MA, 02139, USA
[d]The Institute of Thermomechanics of the Czech Academy of Sciences, Dolejškova 1402/5, Praha 8, ZIP code 182 00, Czech Republic.
[e]Center for Nanoscale Systems, Harvard University, Cambridge, Massachusetts 02138, USA

**Corresponding author:** Maxwell Rae (mraechu@mit.edu)

**Abstract:**

Spinodal decomposition can degrade corrosion resistance and embrittle materials. The ability to quickly, non-destructively detect spinodal decomposition via elastic property measurements before catastrophic materials degradation would represent a significant advance in materials testing. We demonstrate that spinodal decomposition can be detected in a binary Fe-Cr alloy via modulus stiffening using *ex situ* (and *in situ*) transient grating spectroscopy (TGS). The modulus stiffening is attributable to the convex dependence of the $C'$ elastic modulus on chromium content within certain compositional limits, which has been established by prior density functional theory studies. Over this range of initial chromium compositions, chromium partitioning due to spinodal decomposition raises the average $C'$ modulus such that the decomposed Fe-Cr alloy is stiffer than the initial solid solution. The presence and extent of spinodal decomposition in a 36 at. % chromium alloy are confirmed using atom probe tomography (APT), and are consistent with atomistic simulations that predict elastic modulus stiffening in the same alloy. The observed spinodal decomposition via APT is used to predict an expected modulus stiffening which, based on a simple model, agrees in sign and order of magnitude with the observed stiffening via TGS. The quantitative link between an observed composition change and a measured elastic response suggests TGS as a potential tool for non-destructive evaluation of spinodal decomposition in high-chromium ferritic phases in steels.

## Introduction

Spinodal decomposition can limit the service lifetime of critical engineering components. During spinodal decomposition, a single-phase solid solution decomposes into two phases via composition fluctuations that lower the free energy of the system.[1] Though spinodal decomposition is not always detrimental (e.g. spinodal decomposition can strengthen and improve the corrosion resistance of $\beta$-type Ti-24Nb-4Zr-8Sn alloys,[2] and is used to tune the magnetic properties of AlNiCo alloys[3]), the effects of spinodal decomposition can degrade key material properties in Fe-Cr alloys, widely used in power and energy infrastructure. In duplex cast austenitic stainless steel (CASS) alloys, for example, the ductile-to-brittle transition temperature (DBTT) of an in-service component operated at 400°C can shift to room temperature within two months.[4] In addition, spinodal decomposition into $\alpha$ and $\alpha'$ phases within the δ-ferrite phase of the duplex steel causes significant chromium depletion in the $\alpha'$ phase, which degrades corrosion resistance.[5] Currently, no practical non-destructive method exists that directly measures a material property mechanistically linked to spinodal decomposition.

Iron-chromium binary alloys are a useful system to model spinodal decomposition in the chromium enriched δ-ferrite phase of duplex stainless steels (DSS) such as CASS. Embrittlement via spinodal decomposition in the δ-ferrite phase is a well-known life-limiting phenomenon, and literature contains extensive experimental evidence for spinodal decomposition across a range of iron-chromium compositions for temperatures at or below 500°C.

Spinodal decomposition can be characterized at the nanoscale using atom probe tomography (APT) or scanning transmission electron microscopy (STEM)[6–9]. Xiong et al. and Miller et al. performed aging experiments at 500°C across a range of chromium concentrations to identify the chromium composition where the phase transition mechanism changes from nucleation and growth to spinodal decomposition.[10–12] Small-angle neutron scattering (SANS) has also been used to study iron-chromium phase separation.[9,13] At the microscale, hardness measurements can be used to track hardening due to spinodal decomposition.[14] The macroscale effects of spinodal decomposition are often characterized by Charpy impact or fracture toughness testing.[15] In the iron-chromium system, temperature and chromium composition determine the phase transformation mechanism.

At 500°C, the chromium fraction inside the miscibility gap that separates nucleation and growth from spinodal decomposition is disputed.[9–11] SANS and APT indicate nucleation and growth at and below 25 at. % chromium,[9,10] while SANS, STEM, and APT indicate spinodal decomposition above 35 at. % chromium.[6,10,13] Because the miscibility gap widens at lower temperatures, spinodal decomposition extends to lower chromium fractions, which explains its occurrence in CASS alloys. More recently, elastic wave speed measurement via transient grating spectroscopy (TGS) has been proposed as a method to measure spinodal decomposition in duplex steels.[16]

TGS is a laser-based spectroscopic technique in which two pump lasers are crossed at the sample surface to launch two counter-propagating surface acoustic waves (SAWs). The time-dependent intensity of a probe beam reflected off the sample surface is fit to a model to determine the elastic and thermal properties of the material to a micron-scale depth below the illuminated region.[17] For polycrystalline materials with grains larger than the spot size of the TGS probe laser (≈60 μm in this work), the elastic wave speed in a given crystal direction depends on the material's density and elastic stiffness tensor, both of which can vary with temperature and other intensive variables.[18] *Ex situ* TGS rotation studies on single crystals (or large single grains in a polycrystal) enable back-calculation of elastic constants via fitting the angular dispersion of SAW speed measurements, which controls for the effect of crystal orientation on SAW speed.[18] In this work, the periodic composition modulation of spinodal decomposition occurs on the nanometer scale inside of millimeter-scale grains. Because the laser spot size is ≈60 μm, and the surface acoustic wave probe depth of TGS is approximately half the grating spacing (≈3 $\mu$m), the measured elastic response from TGS is effectively from a single grain (a single crystal).[17] The nanoscale composition modulation affects the micrometer-scale surface acoustic response by changing the effective elastic moduli and density of the matrix after spinodal decomposition.

Previous work by Al Dajani et al. showed, via molecular dynamics simulations, that iron-chromium de-mixing—analogous to spinodal decomposition—can stiffen the elastic modulus of iron-chromium alloys.[16] The same study also predicted and measured surface acoustic wave (SAW) and pseudo surface acoustic wave (PSAW) speeds in duplex CASS alloys with and without spinodal decomposition in the δ-ferrite phase. However, the effect of spinodal decomposition on elastic modulus could not be disentangled from material stiffening by other mechanisms (e.g., G-phase precipitation), because CASS alloys are complex engineering materials with multiple alloying elements and a complex microstructure.[16] In addition, the effect of spinodal decomposition on the elastic constants of the medium could not be measured experimentally in CASS, because the crystal orientation relative to the TGS grating was not controlled for, and the δ-ferrite grains in the CASS alloy were smaller than the spot size of the TGS probe laser.[16]

Finally, there is no generally accepted mechanism for why spinodal decomposition stiffens iron-chromium alloys. Non-linearities in the elastic constants of iron-chromium alloys have been shown in two studies using density functional theory.[19,20] The origin of the nonlinearity was studied by Zhang et al. and ascribed in large part to a strong dependence of elastic constants on the magnetic moment in binary iron-chromium alloys, which is itself a strong (non-linear) function of chromium composition.[20] If such non-linearities exist in the elastic modulus as a function of chromium composition, spinodal decomposition could either stiffen or soften the alloy.

This work describes a processing-structure-property relationship for the effect of time at temperature on spinodal decomposition, which stiffens the elastic modulus of an iron-chromium binary model alloy. A mechanism for spinodal stiffening based on the nonlinear dependence of the $C'$ modulus on chromium composition is proposed, and directional and order of magnitude agreement between predicted stiffening (from observed decomposition via APT) and measured stiffening (via TGS) is shown. The presence of spinodal decomposition is measured directly using APT (and indirectly via DSC and hardness measurements). Elastic modulus changes after spinodal are measured using molecular dynamics simulations and inferred using *ex situ* and *in situ* TGS. The binary nature of the iron-chromium system mitigates complicating effects such as G-phase precipitation, which requires additional alloying elements such as nickel. Finally, the value of TGS as a non-contact, non-destructive method to measure spinodal decomposition is discussed.

## Results

### *Ex situ and in-situ TGS detect progressive stiffening in Fe-36Cr alloy*

After the samples were prepared and single grains within each sample were marked, EBSD was performed on the marked region for the Fe-36Cr sample only. The sample preparation method is shown in Fig. 1. EBSD was used to determine the surface orientation of the marked grain. The [$\phi_1$, $\Phi$, $\phi_2$] Euler angles from the Aztec software Kikuchi band fitting algorithm were used to calculate the surface orientation of the razor-outlined grain. For Euler angles of [1.2878, 0.6795, 0.9328] radians, the surface orientation of the razor-marked grain in the Fe-36Cr sample was found to be [0.5048, -0.3743, 0.7779], via a procedure described in the Methods section. The surface orientation was then used to solve the inverse problem to extract changes in elastic constants as a function of time-at-temperature.

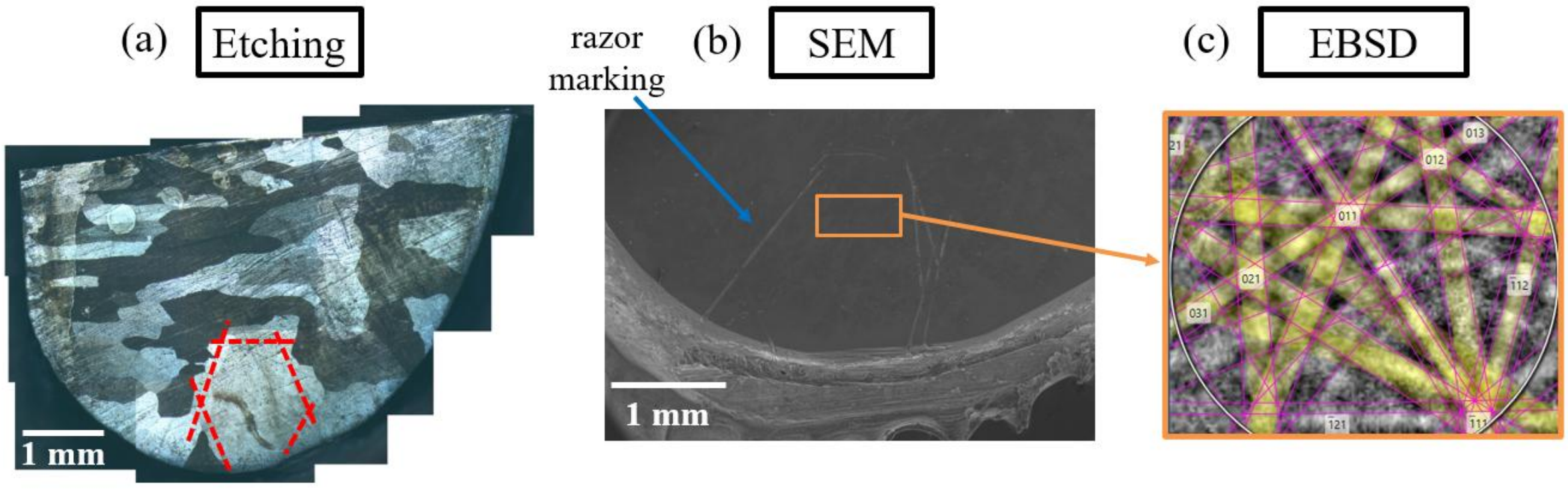

Figure 1. Method of razor marking for successive TGS measurements and determination of grain surface orientation. Shows (a) Fe-36Cr after etching to reveal grain structure (optical image), (b) after razor marking and re-polishing (SEM image), and (c) a representative EBSD pattern. The red dashed outline in the optical image of the etched steel denotes the razor-outlined grain.

TGS results on Fe-36Cr and Fe-26Cr will be presented first, followed by calculation of the change in the elastic constants of Fe-36Cr as a function of heat treatment condition. The various heat treatments are described in the Methods section and summarized in Table 1 below. Table 1 also includes a column for the average SAW speed around a relative minimum of the angular SAW dispersion, where progressive stiffening was most strongly observed in Fe-36Cr. Figure 2 shows the time-temperature history of Fe-36Cr and Fe-26Cr aging, along with the relative minimum SAW speed for Fe-36Cr as a function of heat treatment condition.

The SAW speeds at the relative minimum for Fe-36Cr were measured at 36° relative rotation. The standard deviation in SAW speed measurements in Fig. 2 and 3 were computed using 10 TGS measurements at different areas within a single grain at one relative rotation angle. The uncertainties reported in Fig. 2 and Table 1 were computed from measurements over three nearby orientations, as described in the Table 1 caption. The error in the power spectrum frequency fit used to find the SAW speed was computed, but in every case was found to be approximately two orders of magnitude smaller than the variance in the measurement and was therefore not propagated. The same method was used to compare SAW speeds near a relative minimum (≈36° in Fig. 3) for Fe-26Cr. The angular dispersion of SAW velocities in Fe-36Cr and Fe-26Cr is shown in Fig. 3 for the conditions shown in Table 1.

| **Alloy** | **Heat Treatment** | **Time [hr]** | **Temp. [°C]** | **Min. SAW Speed [m/s]** |
|---|---|---|---|---|
| Fe-36Cr | HOM1 | 2 | 1000 | 2703.6 ± 2.8 |
| | HT1 | 1 | 500 | 2728.7 ± 3.3 |
| | HT2 | 10 | 500 | 2759.4 ± 3.2 |
| | HOM2 | 1 | 1000 | 2649.3 ± 2.8 |
| Fe-26Cr | HOM1 | 2 | 1000 | 2750.6 ± 6.0 |
| | HT1 | 20 | 500 | 2732.2 ± 5.3 |
| | HOM2 | 1 | 1000 | 2735.6 ± 7.5 |

Table 1. Heat treatments and SAW speeds along a soft crystal direction (≈36° for Fe-36Cr, and ≈20° for Fe-26Cr). Errors denote the standard deviation of SAW speeds measured over three orientations (spanning ≈12°) around the relative minimum, which accounts for the ≈1-2° of uncertainty in locating the minimum across aging conditions.

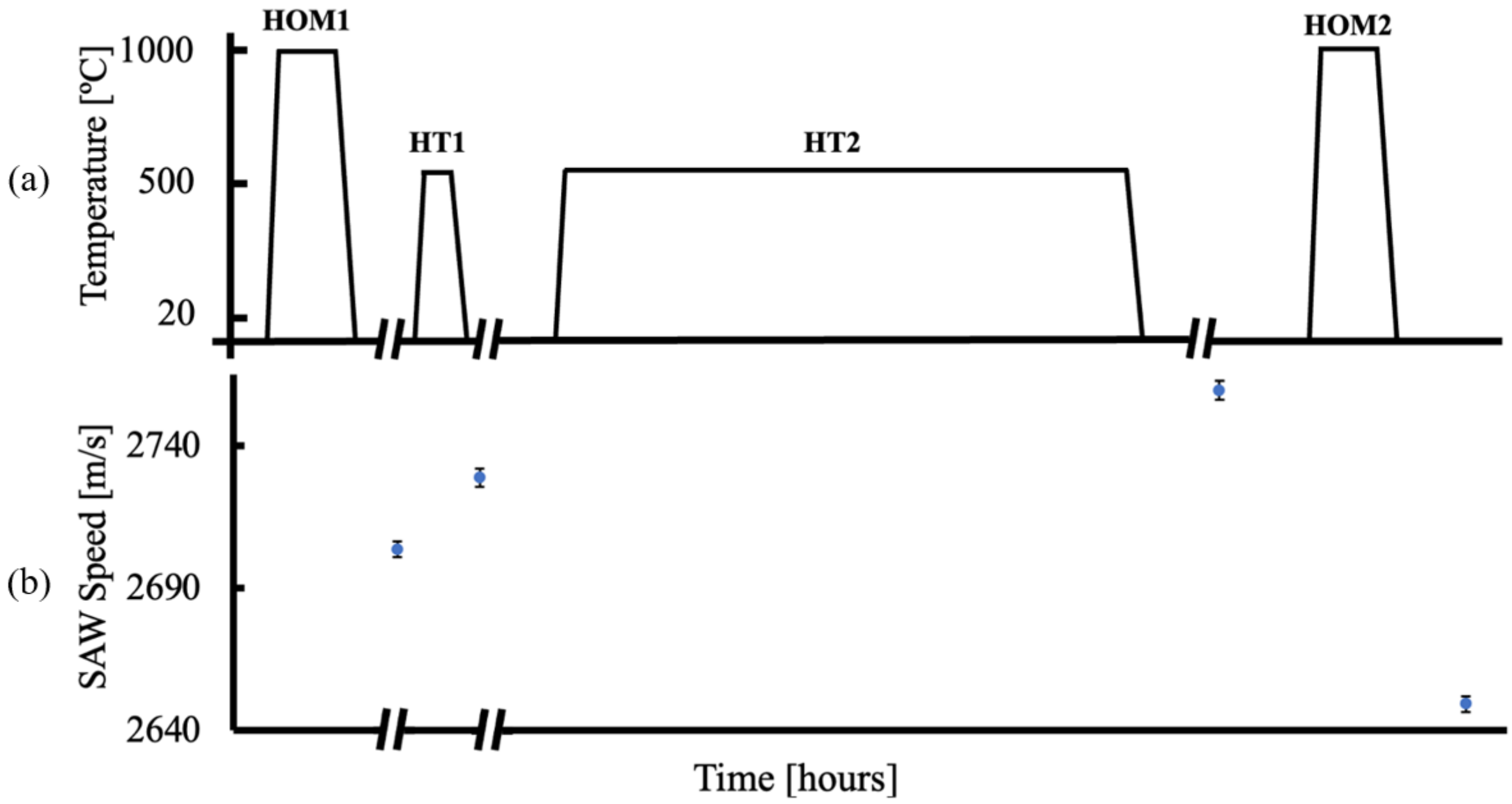

Figure 2. The effect of time at temperature on SAW speed. Shows (a) pictorial representation of time-temperature processing conditions for Fe-26Cr and Fe-36Cr, and (b) SAW speed at relative minimum for Fe-36Cr (error bars are one standard deviation above and below the mean). HOM indicates homogenization and HT stands for heat treatment.

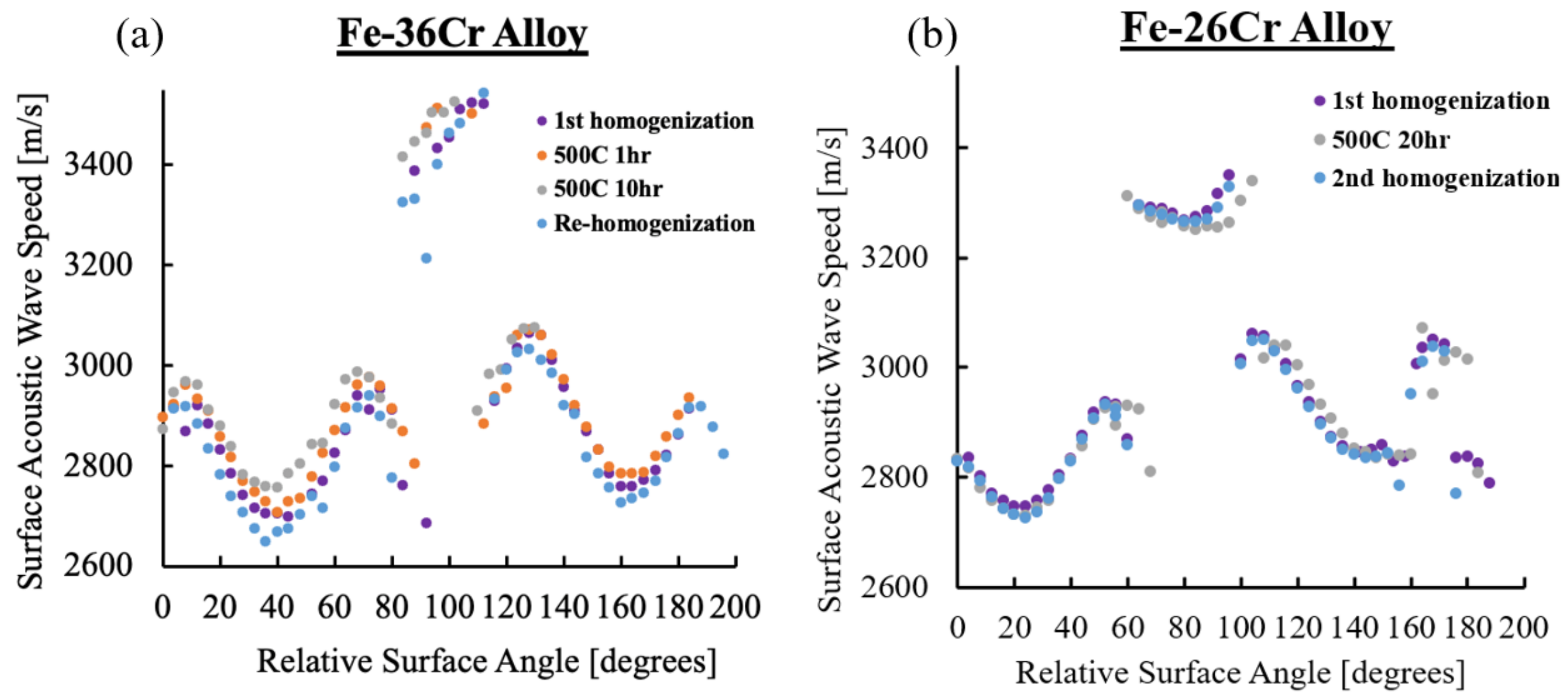

Figure 3. SAW speed as a function of relative surface angle for Fe-36Cr and Fe-26Cr. Each point was taken in the same single grain after each heat treatment, and the standard deviation of SAW speed measurements is ≈0.1% of the measured value, which is too small to see on the plot. Note that stiffening is directionally dependent. Slower directions typically exhibit more stiffening with thermal aging, while faster directions exhibit less, in both (a) the Fe-36Cr alloy, with purple, orange, grey, and blue markers for HOM1, HT1, HT2, and HOM2 conditions, respectively, and (b) the Fe-26Cr alloy with purple, grey, and blue markers for HOM1, HT1, and HOM2, respectively.

The lefthand graph in Fig. 3 shows the angular SAW dispersion for Fe-36Cr for the different heat treatment conditions tested. Using the experimentally measured angular SAW dispersions, the inverse problem was solved to compute the two independent elastic constants that produce significant out-of-plane displacement. The results are shown in Table 2. Notably, $C'$ for both Fe-36Cr samples aged at 500°C, where spinodal decomposition is expected, is more than two standard deviations higher than either homogenized case, where spinodal decomposition is not expected.

Fe-26Cr showed no significant stiffening after aging for 20 hours at 500°C, and in fact showed significant softening (p = 0.0164 for a Welch's test for means comparison).

| **Heat Treatment** | $C'$ **[GPa]** | $C_{44}$ **[GPa]** | **Spinodal expected?** |
|---|---|---|---|
| 1000°C, 2 hr | $56.7 \pm 0.3$ | $113 \pm 0.9$ | No |
| 500°C, 1 hr | $58.2 \pm 0.3$ | $114.7 \pm 0.7$ | Yes |
| 500°C, 10 hr | $59.8 \pm 0.3$ | $111.3 \pm 0.7$ | Yes |
| 1000°C, 1 hr | $55.0 \pm 0.3$ | $112.8 \pm 1.0$ | No |

Table 2. Elastic constants of Fe-36Cr, found by solving the inverse problem. The relationship between $C'$, $C_{11}$, and $C_{12}$ is $C' = 0.5(C_{11} - C_{12})$. Reported error is the 1-sigma fitting error described in ref. [21].

The *ex situ* measurements above required interrupting aging for room-temperature measurements, but TGS can also resolve stiffening *in situ*. *In situ* TGS results were obtained on Fe-36Cr. Two *in situ* tests were conducted at nominal temperatures of 500°C, and 475°C, and in both experiments SAW speed increased as a function of time at temperature. The first *in situ* test (nominal temperature of 500°C) shown in Figure 4 was temperature controlled using an on-sample thermocouple, which resulted in superior temperature control (499.8$\pm$0.5°C) relative to the second test, which was temperature controlled off an in-heater thermocouple. Because the in-heater temperature control was poor and resulted in a temperature rise that confounds SAW measurement data, the results of the second *in situ* test (nominal temperature of 475°C) are provided in the supplementary materials.

Figure 4 shows there is clear stiffening as a function of time at temperatures where spinodal decomposition is expected in the Fe-36Cr alloy. The absolute change in SAW speed during the *in situ* measurement was about 50$\pm$2.1 m/s, or 1.8%, which is much larger than the detection limit of TGS. The *in situ* test also detected a kinetic change approximately ≈60 minutes into the experiment. Before the ≈60-minute mark, the average slope is ≈0.47 m/s/min, and after it decreases to ≈0.09 m/s/min. Given that TGS can detect changes on the order of 0.1% of the mean SAW speed, the slope change should be significant after ≈6 minutes. The results of *in situ* TGS are consistent with ex situ results and demonstrate that TGS is sensitive enough to detect stiffening *in situ* in iron-chromium alloys.

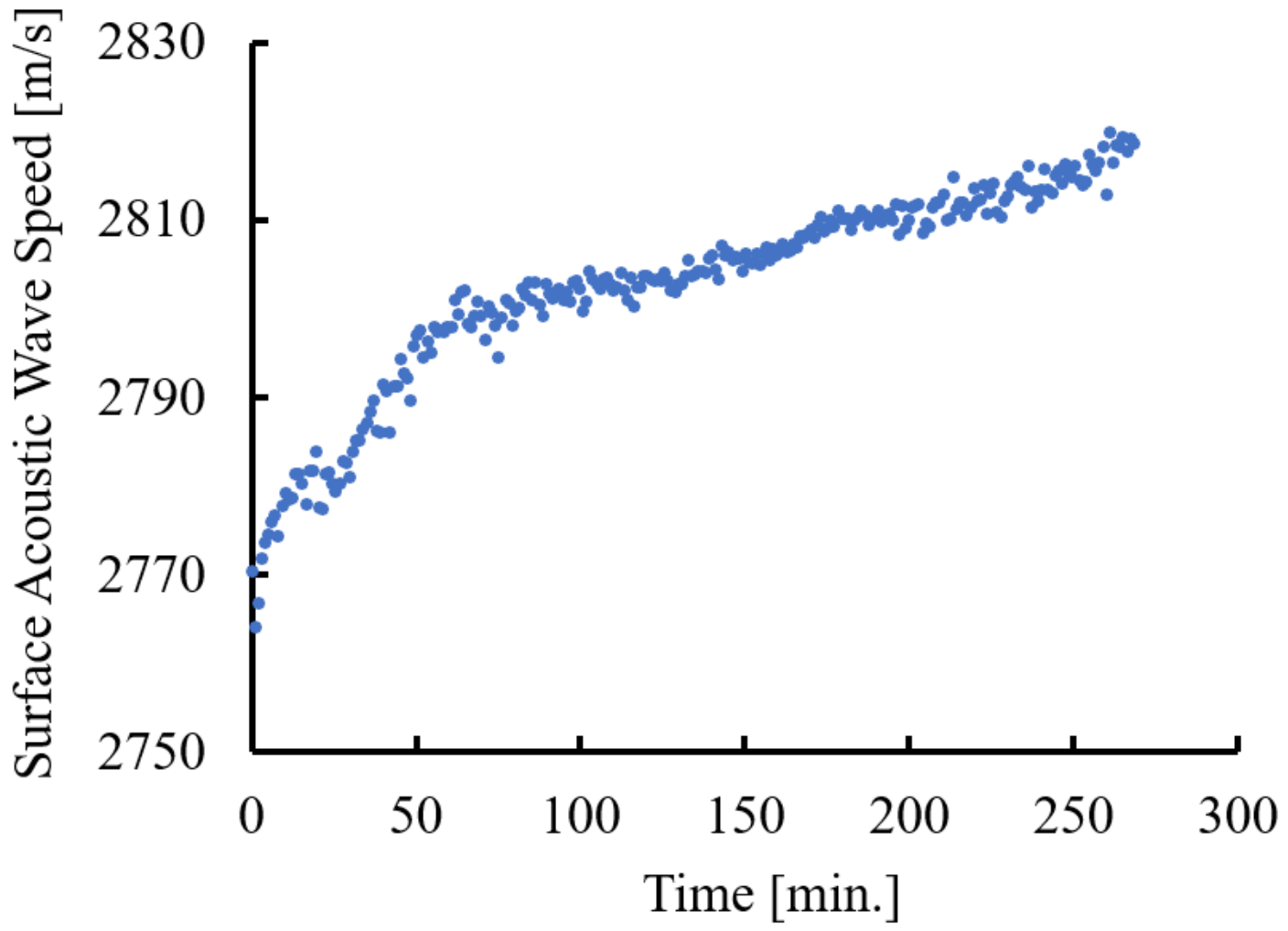

Figure 4. *In situ* TGS at ≈500°C shows clear progressive stiffening. The standard deviation of SAW speed measurements is ≈0.1% of the measured value, which is too small to be seen on the graph.

*APT confirms nano-scale phase separation in aged Fe-36Cr*

APT is the primary method used to determine if spinodal decomposition is present after heat treatment in Fe-36Cr. The purple 38.6 at. % Cr isosurfaces in Fig. 5b show clear phase separation in Fe-36Cr aged for 10 hours at 500°C, and the proximity histogram (proxigram) in Fig. 5c shows the degree of phase separation across the $\alpha/\alpha'$ interface. The proxigram method shows quantitative composition changes across a phase boundary by defining an isoconcentration surface, calculating normal distances of every atom to that surface, then plotting the binned composition in successive distance shells, with negative numbers denoting regions inside the chromium-rich phase.[22]

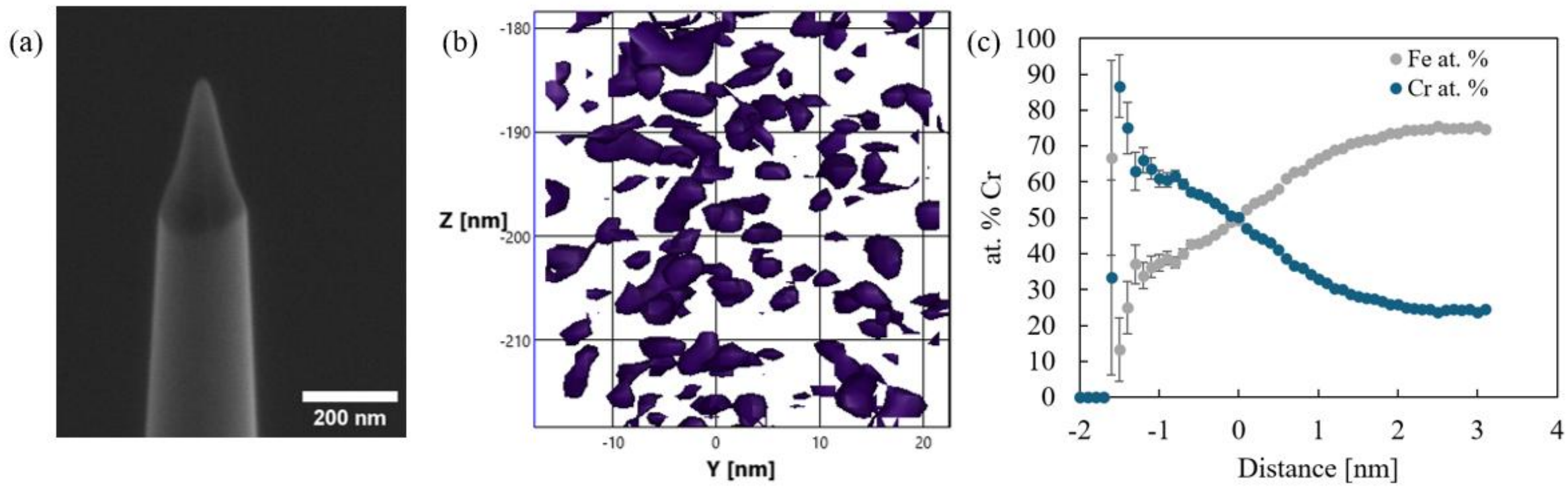

Figure 5. APT reconstruction results on Fe-36Cr aged for 10 hours at 500°C. Shows (a) a secondary electron image of the APT tip, (b) a 2D projection of a 38.6 at. % Cr 3D isosurface from a representative region of the tip, and (c) a proxigram for the same region.

To determine the fraction of the chromium-rich ($\alpha'$) and chromium-poor ($\alpha$) phases after aging, the composition of each phase is measured by APT and the fraction computed via the lever rule. Composition plateaus in the proxigram in Figure 5c show the $\alpha$ and $\alpha'$ chromium fractions (in at. %) are found near $x_{\alpha} = 0.259 \pm 0.002$, and $x_{\alpha'} = 0.617 \pm 0.002$, respectively. The plateau regions were found by measuring the compositions on the chromium-rich and chromium-poor side of the proxigram where the rolling average (n = 5) of the spatial derivative of the chromium and iron at. % are below 5 at. % / nm. The nominal chromium fraction of the alloy is $\bar{x} = 0.363$. Using the lever rule, the phase fractions are $f_{\alpha'} = (\bar{x} - x_{\alpha})/(x_{\alpha'} - x_{\alpha}) = 0.29 \pm 0.004$ and $f_{\alpha} = 0.71 \pm 0.004$, for the chromium-rich and chromium-poor regions, respectively.

As a check, TGS rotation measurements were performed on the grain that APT samples were taken from. The APT samples were from the same Fe-36Cr arc-melted ingot, but in a different grain than the one used for the full TGS rotation study and inverse $C'$ extraction. Identical homogenization and aging were performed on this alloy. EBSD was not performed in the grain where APT was done, so inverse extraction was not attempted. Fig. 6 shows the TGS results, which confirm directionally dependent stiffening in a second grain in the Fe-36Cr sample after 10 hours at 500°C as was observed before in Fig. 3.

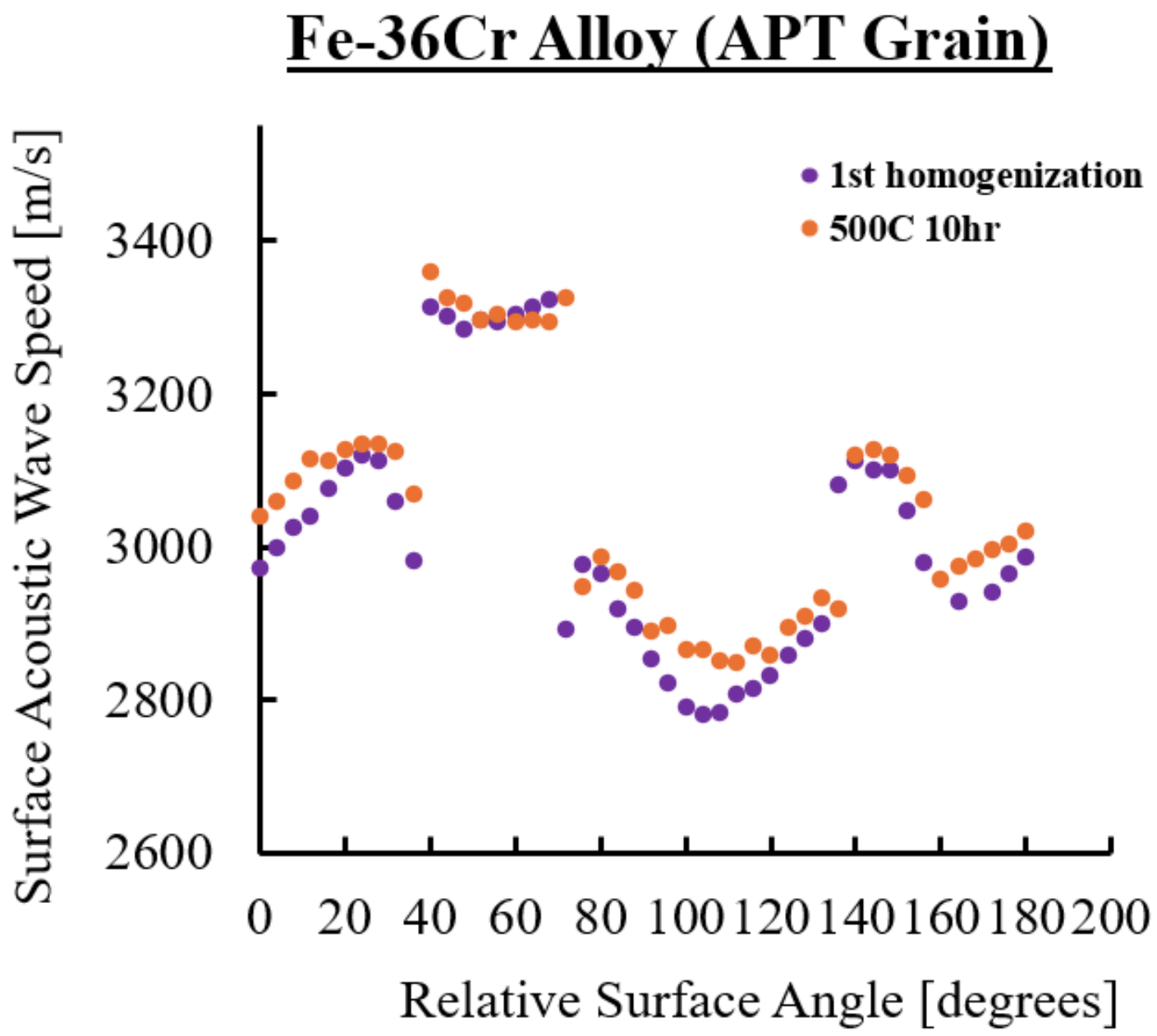

Figure 6. TGS rotation study shows directionally dependent stiffening in the grain where APT was performed. Purple and orange markers are after the 1st homogenization and aging at 500°C for 10 hours, respectively.

After aging at 500°C for 10 hours, the SAW speed rose $2.5 \pm 0.1\%$ along the softest direction (the minimum near 100° in Fig. 6) and a mean of $1.9 \pm 0.1\%$. On average, and preferentially along the softest direction, stiffening is observed by TGS in the same grain that APT showed nanoscale phase separation.

*Hardness testing confirms increase in hardness coincident with increased aging*

As a secondary check for the progressive effect of aging on mechanical properties, hardness testing was performed on Fe-36Cr in the homogenized and as-aged conditions. In the binary iron-chromium model system, progressive hardening is consistent with strengthening driven by spinodal phase separation.[14,23] Fig. 7 shows Fe-36Cr progressively hardens with time at 500°C. Vickers hardness increases from $163 \pm 7$ HV (HOM1 condition) to $221 \pm 11$ HV (500°C, 1hr aged condition), to $284 \pm 8$ HV, where the uncertainty is the standard deviation across at least 9 indents per condition.

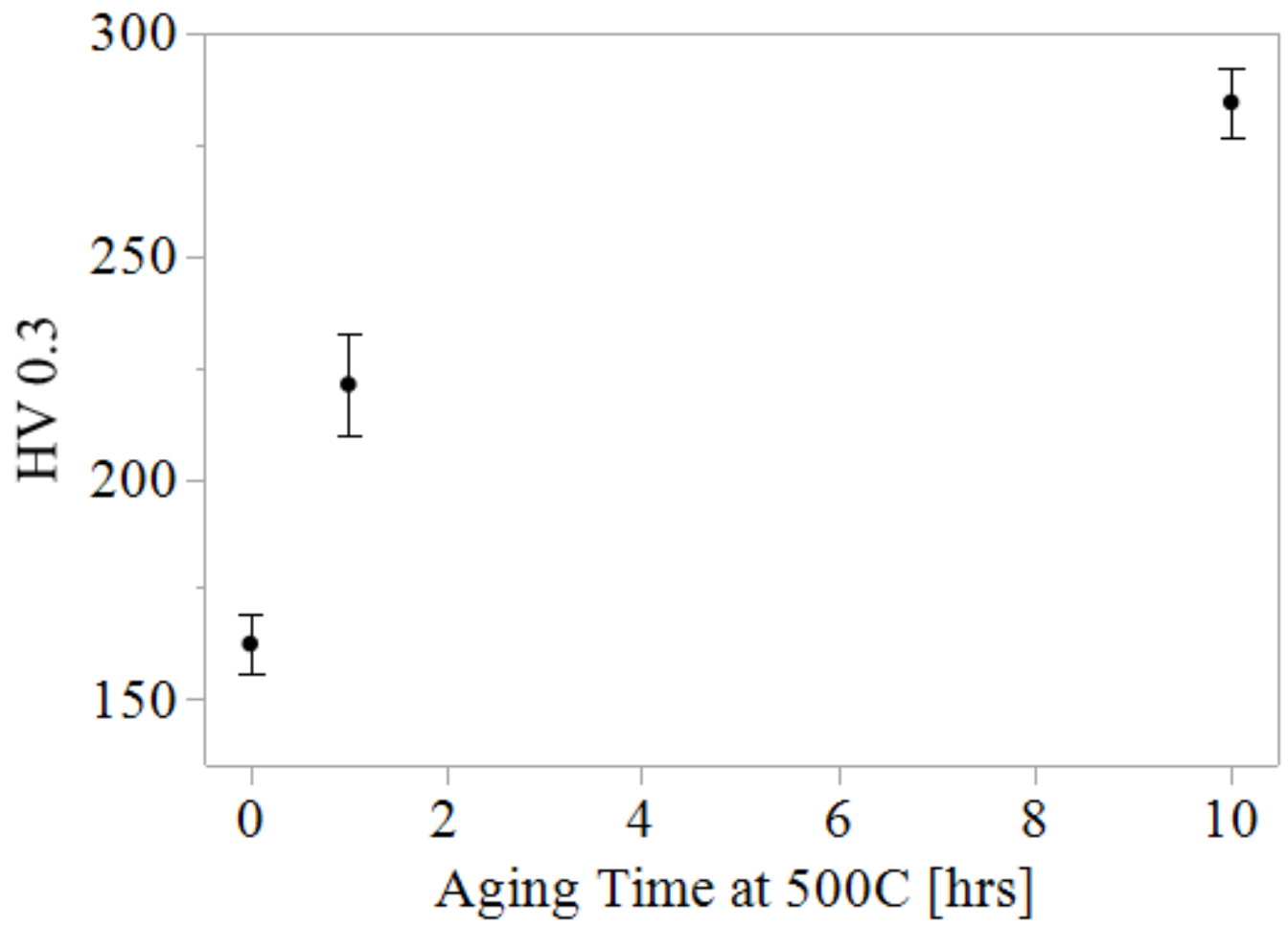

Figure 7. Vickers hardness results as a function of aging time at 500°C for Fe-36Cr. Error bars denote one standard deviation in either direction.

*APT-informed model predicts stiffening of the correct sign and order*

The APT results, in tandem with the expected stiffening due to nonlinearities in the $C'$ modulus as a function of chromium composition, can be used to build a model to predict spinodal stiffening from observed phase separation. Fig. 8 shows empirical and computed values for $C'$ from literature and this study—the key computational dataset of note is the DFT-based exact muffin-tin orbital (EMTO) computational dataset (blue circles),[19] which shows clear non-linearities in the $C'$ modulus as a function of chromium composition, particularly in the region of interest to this study (0.2–0.5 atomic fraction chromium) where spinodal decomposition occurs. Because inverse extraction of elastic constants from TGS is most sensitive to the $C'$ modulus, it was chosen to compare to empirical and computed values from literature.[19,24–28]

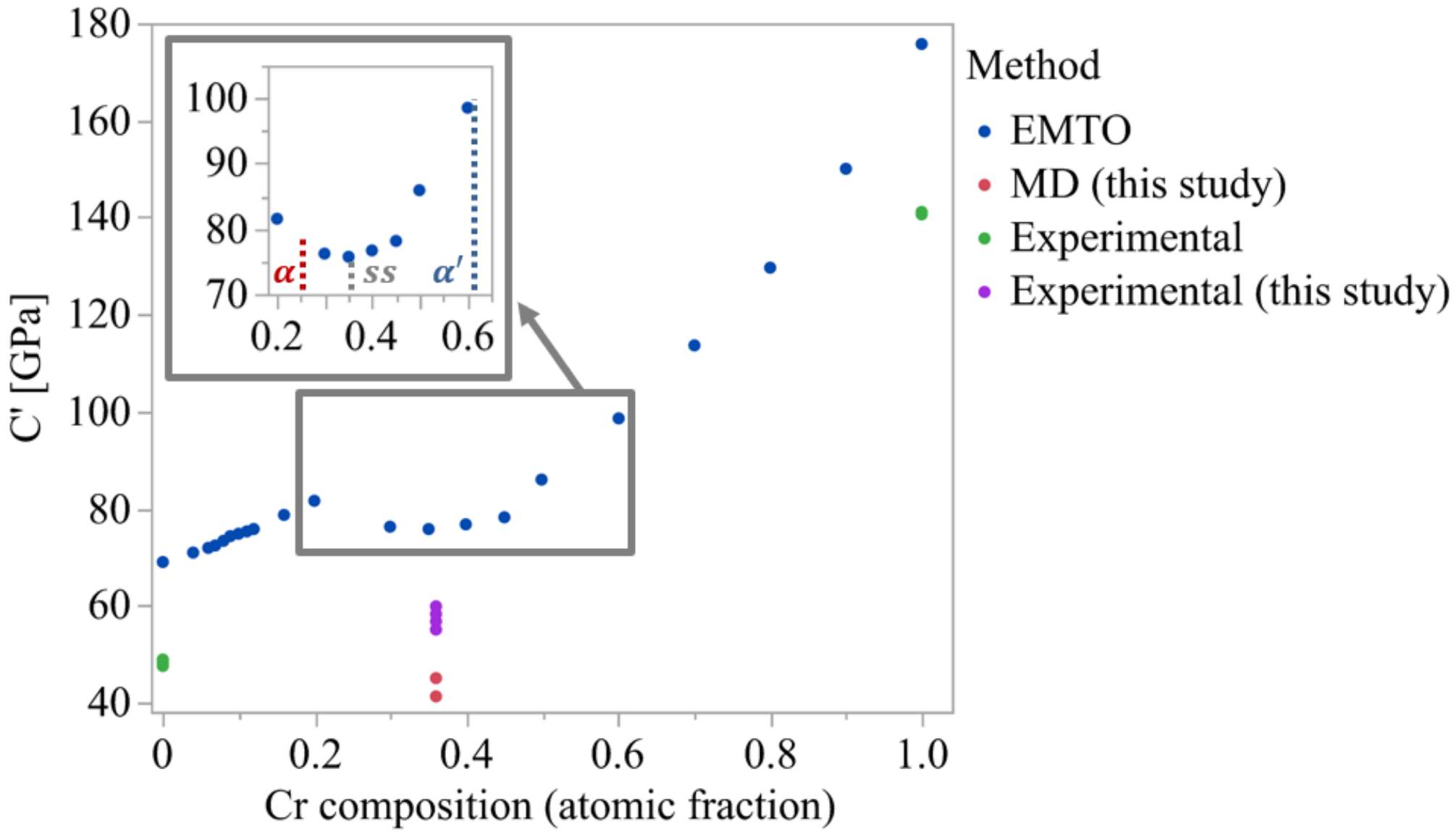

Figure 8. $C'$ modulus as a function of chromium composition. Shown for EMTO (ref. 19) and empirical studies from literature (refs. 24–28) in blue and green markers, respectively, and from MD and TGS inverse extraction in this study, in red and purple markers, respectively.

The region highlighted in orange in Fig. 8 shows the peak and trough values of the chromium composition amplitude for the α and α′ phases of Fe-36Cr aged for 10 hours at 500°C from APT data. Using the APT-measured chromium compositions from the plateau values of the proxigram ($x_\alpha = 0.259 \pm 0.002, x_{\alpha'} = 0.617 \pm 0.002$) and phase fractions ($f_\alpha$ = 0.71, $f_{\alpha'}$ = 0.29), $C'$ values from the EMTO dataset were interpolated at the measured compositions and combined using a simple rule of mixtures approximation.

To predict the effect of phase separation on $C'$ modulus, a simple sinewave model was used for the spatial composition modulation of chromium after spinodal decomposition. The model for chromium composition modulation in space is a 1D sinusoid given by:

$$x(\boldsymbol{r}) = \bar{x} + A \cdot \sin\left(\frac{2\pi}{\lambda}\hat{\boldsymbol{n}} \cdot \boldsymbol{r}\right) \tag{1}$$

Where $\boldsymbol{r}$ is the spatial coordinate in nm, $\bar{x}$ is the total chromium fraction of the alloy, $A$ is the half-amplitude of the composition modulation ($A = 0.5 \cdot [x_{\alpha'} - x_\alpha]$), $\lambda$ is the spinodal wavelength in nm, and $\hat{\boldsymbol{n}}$ is the modulation direction. For isotropic spinodal, the expression simplifies to:

$$x(r) = \bar{x} + A \cdot \sin\left(\frac{2\pi}{\lambda}r\right) \tag{2}$$

Using the proxigram data for $A$, the spatial composition modulations of spinodal decomposition can be used to estimate changes in average $C'$ modulus, which can be directly compared to $C'$ modulus changes measured by TGS for example by a rule-of-mixtures approach such that:

$$C'_{avg} = f_\alpha C'_\alpha + f_{\alpha'} C'_{\alpha'} \tag{3}$$

Where $C'_{avg}$ is the average $C'$ modulus, and the $C'$ moduli of the $\alpha$ and $\alpha'$ phases are $C'_\alpha$ and $C'_{\alpha'}$, respectively. Notice the spinodal wavelength does not enter into this model, but would need to be accounted for in a model that takes coherency effects into account.

Using the above model, the $C'$ modulus computed for the aged alloy was 81.3 GPa, compared to 76 GPa for the nominal Fe-36Cr alloy. The estimated increase of $C'$ by 7.1 ± 0.2% after spinodal decomposition can be compared to 5.4 ± 0.5% measured experimentally by TGS. These values agree in sign and order of magnitude. The error in the sine model $C'$ increase is calculated by propagating the APT composition error, and does not include systematic error, which may come from the EMTO data extrapolation to finite temperature or the amplitude plateau estimation from the proxigram.

*Atomistic simulations also predict elastic stiffening upon spinodal decomposition*

Hybrid MD/MC simulations were performed to support the experimental analysis. Fig. 9 shows the Fe-26Cr and Fe-36Cr alloys before and after thermal equilibration, revealing clear phase separation in both cases into Fe-rich and Cr-rich regions. The approach described in ref. 7 was used to calculate the spinodal decomposition wavelength and amplitude for both alloys in the thermally equilibrated state. Because the distributions of both spinodal wavelength and amplitude are skewed, the median and interquartile ranges of each are reported. The median spinodal wavelength for the Fe-26Cr and Fe-36Cr alloys were 2.75 nm and 2.89 nm, respectively, with 25$^{th}$ and 75$^{th}$ percentile ranges of 1.92-3.98 nm and 2.06-4.13 nm. The median composition amplitude (mole fraction) for Fe-26Cr and Fe-36Cr was 0.18 and 0.26, respectively, with 25$^{th}$-75$^{th}$ percentile ranges of 0.14-0.22 and 0.21-0.32.

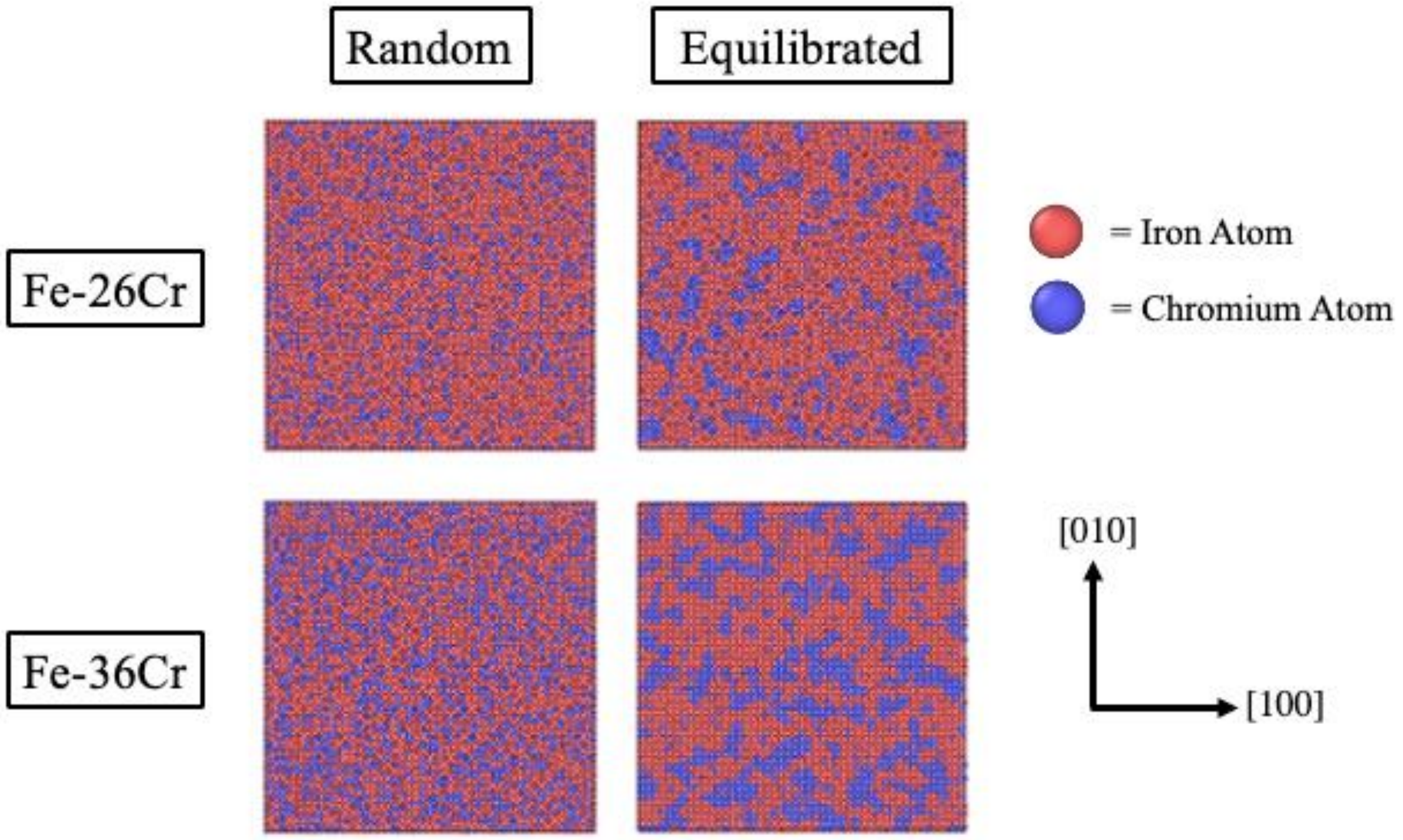

Figure 9. OVITO visualizations of Fe-26Cr and Fe-36Cr in solid solution and thermally equilibrated (at 500°C) states used for elastic constant calculations. Red voxels denote iron atoms, and blue voxels denote chromium atoms, and the black arrows denote crystallographic direction.

Molecular dynamics was also used to examine changes in material properties before and after thermal equilibration at 500°C. Specifically, the elastic stiffness tensor and density were measured for both states:

$$C_{ij}^{ss} = \begin{bmatrix} C_{11} & C_{12} & 0 \\ C_{12} & C_{11} & 0 \\ 0 & 0 & C_{44} \end{bmatrix} = \begin{bmatrix} 184.77 \pm 0.13 & 102.32 \pm 0.1 & 0 \\ 102.28 \pm 0.1 & 184.77 \pm 0.14 & 0 \\ 0 & 0 & 85.52 \pm 0.05 \end{bmatrix} \quad 4$$

$$C_{ij}^{aged} = \begin{bmatrix} 195.4 \pm 0.2 & 105.42 \pm 0.13 & 0 \\ 105.41 \pm 0.13 & 195.39 \pm 0.21 & 0 \\ 0 & 0 & 89.2 \pm 0.05 \end{bmatrix} \quad 5$$

Where $C_{ij}^{ss}$ and $C_{ij}^{aged}$ represent the elastic stiffness tensors for the solid solution and thermally equilibrated systems, respectively, in Fe-36Cr. The units for all elastic stiffness tensor components are in GPa. On a percentage basis, the average stiffness increases for $C_{11}$, $C_{12}$, and $C_{44}$, were 5.74 $\pm$ 0.07%, 3.03 $\pm$ 0.09%, and 4.32 $\pm$ 0.04%, respectively. Errors represent one standard deviation about the mean, computed from 10 independent simulations for each alloy before and after thermal equilibration, with errors summed in quadrature. In addition to the elastic stiffness tensor, the density of Fe-36Cr in the solid solution and aged conditions was computed via MD and found to be 7367 $\pm$ 2 kg/m$^3$ and 7444 $\pm$ 2 kg/m$^3$.

For Fe-26Cr, the average stiffness increases for $C_{11}$, $C_{12}$, and $C_{44}$, were 3.84 $\pm$ 0.14%, 2.52 $\pm$ 0.14%, and 3.28 $\pm$ 0.08%, respectively. The density of Fe-26Cr in the solid solution and aged conditions were computed to be 7483$\pm$2 kg/m$^3$ and 7531 $\pm$ 2 kg/m$^3$.

## Discussion

To establish a direct link between spinodal decomposition and stiffening, evidence was presented for both effects. That spinodal decomposition occurs in Fe-36Cr alloys at 500°C is well established in the literature[6,10,13]—APT was used for direct verification of phase separation in the sample used for *ex situ* TGS studies. In Fe-36Cr, a progressive SAW speed increase after aging was observed, and APT confirmed nanoscale phase separation into Cr-rich and Cr-poor regions after aging Fe-36Cr for 10 hours at 500°C. DSC results in the supplemental information provide secondary evidence for phase separation based on a positive unmixing enthalpy measured in the 500°C, 10-hour aged Fe-36Cr sample. Furthermore, SAW stiffening was directly measured in the same grain where APT was performed. Progressive hardening of Fe-36Cr with time at 500°C was also observed, and is consistent with strengthening driven by spinodal phase separation[14]. The results presented establish that both spinodal decomposition and stiffening occurred in the aged Fe-36Cr alloy. To close the gap between microstructural characterization of phase separation and direct measurement of SAW stiffening, an APT-informed model based on nonlinear $C'$ modulus changes was presented and showed good agreement between $C'$ modulus stiffening of Fe-36Cr after aging via TGS (5.4 ± 0.5%) and via the APT-informed model (7.1 ± 0.2%). Next, we turn to why spinodal produced stiffening in this specific alloy system.

Mechanistically, stiffening (or softening) of an alloy can arise from more than one source. Here we focus on the contribution directly testable with our data, namely the nonlinear dependence of the elastic constants on composition. The effect of spinodal decomposition is separation into two phases, α and α′, which may have different elastic moduli from their parent phase, and from each other. To first order, a rule of mixtures approximation of the change in $C'$ modulus due to complete phase separation can be computed. However, nonlinearities in the elastic constants of iron-chromium alloys have been shown in two studies using DFT, which allow direct predictions based on measured phase separation via APT.[19,20] The origin of the non-linearities was studied by Zhang et al. and ascribed in large part to a strong dependence of elastic constants on the magnetic moment in binary iron-chromium alloys,[20] which is a strong (nonlinear) function of chromium composition. Two areas of note in Fig. 8 are the relative maximum in $C'$ centered around 20 at. % chromium, where spinodal decomposition would be expected to soften the $C'$ modulus, and the relative minimum in $C'$ centered around 35 at. %, where spinodal would be expected to stiffen the elastic modulus. Both regions are relevant to this study.

The APT-informed rule of mixtures estimate of 11.8% stiffening in the aged Fe-36Cr alloy is an upper bound on the $C'$ modulus increase because the calculation assumes complete phase separation into α and α′ phases (i.e. an abrupt composition change at the α/α′ interface), when the composition change across an interface is continuous. The smoothed sinewave estimate of 7.1 ± 0.2% change in $C'$accounts for diffuse interfaces and is in good directional and order-of-magnitude agreement with the 5.4 ± 0.5% measured experimentally by TGS. Compared to the as-homogenized state, heat treatments of 1 and 10 hours at 500°C result in anisotropic stiffening, with greater stiffening along initially elastically soft directions (local minimum in Fig. 3), as shown in Table 2. Spinodal decomposition is known to modulate along elastically soft directions in cubic alloys with negative elastic anisotropy to minimize elastic strain energy.[29] This explains why in Figs. 3a and 6, we see the strongest stiffening upon spinodal decomposition in the softest directions, or those with the slowest initial SAW speeds.

In Fig. 8, the $C'$ modulus calculated by EMTO is systematically higher than $C'$ measured experimentally. The EMTO elastic constants are static 0 K results that neglect explicit phonons and finite temperature magnetic excitations. Thermal expansion, anharmonic phonons, longitudinal spin fluctuations, magnetic short-range order with non collinear moments, and dynamic spin-lattice coupling all lower $C'$ at 298–300 K, so the EMTO $C'$ values are higher than experiments, as expected.[20,30–34] In terms of percentage differences, however, the $C'$ modulus changes predicted empirically and by the sine-wave model agree well. The APT-informed model based on EMTO predictions of $C'$ may overpredict stiffening because EMTO values are static 0 K values, and the $C'$ dependence on chromium composition may flatten at finite temperature. Alternatively, the proxigram amplitude from APT may overstate the volume-averaged composition modulation, which would in turn overpredict the $C'$ modulus increase. Finally, the weaker composition dependence of $C'$ in the MD/MC simulations compared to EMTO likely reflects that the embedded-atom method potential captures magnetism only implicitly, rather than reproducing the composition dependent magnetoelastic structure of $C'(x_{cr})$.[20,35]

Beyond the agreement between the APT-informed EMTO predictions and empirically measured $C'$ modulus, there is reproducible quantitative agreement between the two TGS rotation studies done on different grains in Fe-36Cr. The directly measured $C'$ modulus stiffening on the EBSD-characterized grain is 5.4 ± 0.5%. In the grain where APT was performed, the SAW speed in the softest direction rose by 2.5 ± 0.1%, and on average 1.9 ± 0.1% across the entire SAW dispersion. Because SAW speed is proportional to the square root of the modulus over the density ($v \propto \sqrt{C/\rho}$, where $v$ is the SAW speed, $E$ is crystal direction dependent elastic modulus, and $\rho$ is density), we can convert the SAW stiffening to expected relative modulus stiffening with an assumption on the density change after aging. MD/MC predicted a 1.05% density increase after aging—if we use this value, the predicted average expected modulus change after aging in the APT grain is 4.7 ± 0.3%, and 5.9 ± 0.3% along the soft direction. The TGS rotation study results on the APT grain are in quantitative agreement with the results in the EBSD-characterized grain.

By converting the elastic constants $C_{11}$ and $C_{12}$ into $C'$ via $C' = 0.5(C_{11} - C_{12})$, and directly comparing $C_{44}$, the relative changes in elastic constants predicted by hybrid MD/MC and those measured by solving the inverse problem can be compared. However, before direct comparison, discrepancies in absolute values must be addressed. For example, the 1000°C, 2 hr treated Fe-36Cr sample in Table 2 has a $C'$ modulus of $56.7 \pm 0.3$, while the solid solution Fe-36Cr $C'$ modulus from MD/MC was found to be $41.24 \pm 0.16$. In absolute terms, the values are not in close agreement, but relative changes due to spinodal decomposition are more significant in this context. Comparing relative rather than absolute values reflects that interatomic potentials in atomistic simulations typically capture trends in material response more reliably than exact physical constants. Although absolute elastic constants depend strongly on the potential's parameterization, relative changes across thermodynamic states—such as before and after spinodal decomposition—can consistently represent correct physical behavior. The inverse extraction of the $C'$ elastic constant is the most precisely determined of the three independent elastic constants because the value of $C'$ is strongly correlated with surface acoustic waves. In contrast, $C_{44}$ depends more strongly on pseudo-surface acoustic waves, which are more difficult to evaluate from the experimental data taken in this study.

From hybrid MD/MC, the relative change in $C'$ from random solid solution to the thermally equilibrated state at 500°C is 9.1 ± 0.3%, where the uncertainty from MD/MC is the standard deviation of the ten elastic constant measurements taken before and after thermal equilibration. The $C'$ modulus for the 1- and 10-hour aged samples increases at a decreasing rate from 2.6 ± 0.5% to 5.4 ± 0.5%, respectively, which is below the change predicted by MD/MC after full thermal equilibration at 500°C for Fe-36Cr. The MD/MC thermally equilibrated results represent a long-term aging condition that vastly exceeds the experimental maximum aging time of 10 hours, but there is reasonable agreement between hybrid MD/MC predictions and experimental results, which provides further evidence for both the magnitude and direction of the stiffening caused by spinodal decomposition in Fe-36Cr aged at 500°C.

Fe-26Cr showed a small but significant SAW softening of -0.7 ± 0.3% on average after aging for 20 hours at 500°C. Because the magnitude of the SAW decrease is small, and no characterization was done on Fe-26Cr, it is difficult to make the same processing-structure-property connection shown for Fe-36Cr. The small magnitude of SAW speed decrease is consistent with the Fe-26Cr being in a more linear region of the $C'(x_{cr})$ curve, where small composition modulations have minor effects on the $C'$ modulus. Hybrid MD/MC on Fe-26Cr predicts $C'$ stiffening of 5.5 ± 0.3% after phase separation via thermal equilibration, which is opposite the measured modulus change in Fe-26Cr. This apparent discrepancy is consistent with the embedded-atom method potential capturing the thermodynamics of spinodal decomposition while missing the composition dependent magneto-elastic structure of $C'(x_{cr})$ responsible for the softening (or, more precisely, lack of stiffening) observed.

In addition to *ex situ* TGS data, the results of *in situ* TGS are presented in part in Fig. 4 (and in full in the supplemental information). The *in situ* test at a nominal temperature of 500°C shows clear progressive stiffening in Fe-36Cr, which is consistent with *ex situ* results and demonstrates that TGS can resolve stiffening in real time. In both *in situ* experiments the stiffening rate dropped significantly after one hour, which implies a kinetic transition on a similar timescale to that observed by Hörnqvist et al., who performed *in situ* SANS on an Fe-36.9Cr at. % alloy for 10 hours at 500°C.[13] After ≈1 hour at 500°C, Hörnqvist et al. observed the characteristic spinodal length scale shift from oscillatory behavior to the expected Cahn-Hilliard-Cook power-law growth.[13] Although TGS measures an elastic response at the mesoscale and SANS probes nanoscale composition fluctuations, the shared kinetic transition observed at ≈1 hour demonstrates TGS is sensitive enough to track small changes in the progression of spinodal decomposition in Fe-36Cr.

The binary nature of the Fe-Cr alloys in this study eliminates competing effects observed in more complex CASS alloys, particularly G-phase precipitation, and shows experimental agreement with a proposed mechanism for spinodal stiffening, namely nonlinear changes in the $C'$ modulus as a function of chromium composition. TGS was shown to be sufficiently sensitive to measure SAW speed changes in both Fe-36Cr and Fe-26Cr under the time-at-temperature aging conditions used in this study. In general, SAW speed changes on the order of ≈5 m/s (0.1%) are measurable. TGS is a relatively inexpensive, non-destructive technique capable of accurately tracking elastic properties both *in situ* and *ex situ*. Developing a commercial TGS system for *in operando* monitoring of ferritic or duplex steels would require integrating recent advances in TGS mapping and miniaturization, along with carefully controlled component tests over extended time periods.[36,37]

These results demonstrate that TGS can sensitively track the elastic response associated with spinodal decomposition in the Fe-Cr system. Translating this approach to complex engineering alloys like CASS, or to other systems where coherency effects must be accounted for, will require separating the contribution of each mechanism to the TGS-measured elastic response, through modeling or empirical validation verified by microscopy or other complementary methods. Future TGS monitoring systems must also account for local grain orientation effects, either by carefully repeating each successive measurement on the same area and with the same grating orientation, or by measuring the angular SAW dispersion at fixed locations on the in-service component. Within those constraints, *in situ* and *ex situ* TGS measurements may provide a practical route toward non-destructive monitoring of thermal aging.

## Methods

*Sample Preparation and Aging Conditions*

The Fe-36Cr sample (see Table 1) was prepared using a Centorr arc melter with a tungsten electrode under an argon environment. The chamber was evacuated and flushed with 99.999% pure argon 10 times prior to each melt, and the alloy was flipped and re-melted 3 times to improve compositional homogeneity. A ≈20-gram sample was made with 99.99% pure iron pieces and 99.995% pure chromium pieces. Fe-26Cr came from a historical sample (method of production unknown), but subsequent high temperature homogenization and composition measurements enable Fe-26Cr to be reproduced by any number of methods. Samples were sectioned with a diamond saw to ≈2 mm thick, and ≈ 100 mL of Kalling's reagent number 2 was used to etch the sample for ≈90 seconds, which revealed the grain microstructure. A grain with a minimum dimension larger than 200 μm was selected in the Fe-36Cr and Fe-26Cr alloys by marking with a razor blade. Selective grain marking was done on the sample stage of a Zeiss ML-100 optical microscope, to facilitate identification of a large grain for *ex situ* TGS studies.

Alloy heat treatments were done using an MTI KSL-1200X annealing furnace. Samples were drop-quenched in deionized (DI) water after each heat treatment, and any oxidation was polished off the sample before TGS measurement using 800 and 1200 grit SiC sandpaper, then 3-, 1-, 0.25-, and 0.1-μm diamond solutions, followed by a 0.05-μm OPS alumina polish. Homogenization and subsequent heat treatments were done in air in a KSL-1200X muffle furnace with 0.1°C temperature control accuracy achieved using a Eurotherm temperature controller. Both alloys underwent the same homogenization treatment at 1000 $\pm$ 0.1°C in air for 2 hours, followed by a drop quench in DI water. Subsequent heat treatment procedures were performed at 500 $\pm$ 0.1°C, also followed by drop quenching in DI water. A final re-homogenization was performed for 1 hour at 1000 $\pm$ 0.1°C, followed by a drop quench in DI water. After the initial homogenization, and in between each heat treatment, TGS measurements were taken. Alloy compositions were measured by inductively coupled plasma optical emission spectroscopy (ICP-OES) with a detection limit of 100 parts per million for trace metals, results for this analysis are shown in Table 3.

| Alloy Name | Fe [at. %] | Cr [at. %] | Trace Elements [at. %] |
|---|---|---|---|
| Fe-26Cr | 73.9 | 26.1 | 0.012 (Ti), <0.01 (All other) |
| Fe-36Cr | 63.7 | 36.3 | <0.01 (All) |

Table 3. Alloy compositions measured by ICP-OES with ASTM E 1479-16.

*Electron Backscatter Diffraction Measurements*

Electron backscatter diffraction (EBSD) measurements were done on a JEOL JSM 7900F Schottky FE-SEM at 20 keV using an Oxford Symmetry EBSD detector. Aztec EBSD software was used to extract the Euler angles of the grain that was marked and used for TGS studies. The Euler angles relate the coordinate system of the SEM image to the crystal coordinate system of the material. A computer code was written and used to extract the crystal surface orientation of the razor-marked grain from the set of Euler angles given by EBSD using the appropriate rotation matrix. The code used to perform this matrix algebra can be found in the data availability GitHub repository for this paper.[38]

*Transient Grating Spectroscopy and Inverse Elastic Constant Calculations*

Two TGS instruments were used, one for *ex situ* measurements at room temperature (in between heat treatments), and one for separate *in situ* experiments. *Ex situ* TGS measurements were done on a system described in ref. 16. All TGS measurements were taken at a nominal grating spacing of 6.4 μm—exact grating spacings for each experiment are indicated in the data availability repository.[38] Elastic wave speeds were determined by taking the power spectrum of the raw TGS trace, and fitting a Lorentzian to the resulting spectrum.

A subset of the independent elastic constants can be calculated by solving the so-called "inverse problem" using the SAW speed as a function of rotation angle.[18] To solve the inverse problem and compute changes in elastic constants, the sample must be a single grain with dimensions larger than the spot size of the TGS probe laser (≈60 μm 1 $\sigma$ width) and a known surface orientation. The surface orientation of the crystal determines which crystal directions the laser-induced grating probes, and thus the elastic wave speed as a function of the physical sample orientation angle. In this work, the surface orientation of the grain used in *ex situ* TGS studies was determined via EBSD. Then, using the mass density and the surface orientation of the sample, a computer code iteratively tested different sets of elastic constants to find those that minimize the sum-square deviation of the angular dispersion of SAW velocities from the measured dispersion.[21] A detailed description of the approach is provided in ref. 21. The TGS parameters used in this work are reported in Table 4.

| Parameter | Value |
|---|---|
| Calibrated on-sample grating spacing | 6.4 μm |
| Pump wavelength | 532 nm |
| Pump energy per pulse | ≈6 μJ |
| Pump pulse duration | <500 ps |
| Pump spot size (on-sample) | ≈140 μm |
| Pump firing frequency | 1 kHz |
| Probe wavelength | 785 nm |
| Probe frequency | continuous wave |
| Probe spot size | ≈105 μm |
| Detector bandwidth | 50 kHz – 1 GHz (3 dB) |
| Oscilloscope bandwidth | 5 GHz |
| Signal trace averages | 10,000 |

Table 4. TGS parameters used for *ex situ* work.

*In situ* TGS measurements were done using a TGS system described in ref. 39, *with in situ* heating under $< 1 \times 10^{-5}$ Torr vacuum using a HeatWave button heater (catalog number #101275-28) and continuously monitored temperature via either an in-heater or on-sample K-type thermocouple. Both *in situ* measurements were done using a nominal grating spacing of 6.4 μm. In the two *in situ* experiments performed (one presented in the main manuscript, and one provided in the supplemental information), two different temperature control methods were used. In the first experiment, presented in the main manuscript, a K-type thermocouple was spot-welded to the surface of the Fe-36Cr alloy, and temperature was controlled using the on-sample thermocouple. In the second in-situ experiment, an on-sample thermocouple was used for continuous temperature monitoring, and an in-heater K-type thermocouple was used for temperature control (both K-type thermocouples).

Temperature control using the on-sample thermocouple resulted in temperature stability within 1°C once 500°C was reached (499.8±0.5°C). By contrast, when temperature was controlled by the in-heater thermocouple, the on-sample temperature continued to rise over the course of the experiment at a rate of ~1.5°C/hour (from 473.1°C to 475.8°C during the first 1.7 hours). The nominal temperature for the in-heater TC control sample was 475°C.

*Atom probe tomography*

Atom probe tomography was performed on a Cameca LEAP 4000X HR at a base temperature of 40 K, 80 pJ pulse energy, a pulse rate of 100 kHz, a detection rate of 1%, and a 50 ps pulsed laser (355 nm wavelength). The reconstruction and proxigram used for quantitative analysis were compiled using the Cameca AP Suite 6 software.

*Hardness measurements*

Vickers microhardness measurements were done on a Struers/Emco-Test DuraScan 70 Vickers Hardness Tester and conformed with standards EN ISO 6507 and ASTM E384. A load of 300 g (2.94 N) was used with a dwell time of 10 seconds, and 9-11 measurements per condition.

*Atomistic simulations*

Hybrid molecular dynamics/Monte Carlo (MD/MC) simulations were performed using the Large-scale Atomic/Molecular Massively Parallel Simulator (LAMMPS)[40] to thermally equilibrate the Fe-Cr system, effectively replicating the heat treatment process conducted experimentally. A body-centered cubic (BCC) cell was first initialized by randomly distributing Fe and Cr atoms according to either the 26 or 36 at. % Cr composition. The system was oriented such that the <100> crystallographic directions aligned with the x, y, and z axes of the simulation cell. The simulation box dimensions were 13.6 nm × 13.6 nm × 13.6 nm, containing 207,646 atoms. These dimensions were chosen to be sufficiently large to accommodate the experimentally observed spinodal decomposition wavelength of 4.5 nm[9]. Periodic boundary conditions were applied in all three directions, and atomic interactions were modeled using an embedded-atom method (EAM) interatomic potential[35].

The system was initially relaxed at 773 K and zero pressure using a Nosé–Hoover thermostat and barostat, both with a damping parameter of 1 ps and chain length of 3. Subsequently, MC swap attempts were introduced between atoms of different elements, with swaps accepted or rejected based on the Metropolis criterion[41,42]. This probabilistic rule accepts swaps that lower the system energy or, alternatively, accepts energy-increasing swaps with a probability of $\exp(-\Delta E/k_B T)$, where $\Delta E$ is the change in system energy, $k_B$ is the Boltzmann constant, and $T$ is the temperature (set to 773 K). Each MC step consisted of 10 attempted atom swaps, followed by 10 ps of MD relaxation at 773 K. A timestep of 1 fs was used for the MD steps. A total of 200,000 MC steps were performed to ensure full equilibration of the structure (equivalent to a total of 10 MC steps per atom). To ensure statistical robustness, the simulation was repeated 10 times with different initial random configurations.

The elastic constants of both the random and thermally equilibrated systems were evaluated by measuring changes in the average stress tensor in response to finite deformations of the simulation cell volume[43] at 300 K. Elastic constants $C_{11}$, $C_{12}$, and $C_{44}$ represent the computed components of the stiffness tensor in Voigt notation. All atomistic simulation visualizations were performed using the Open Visualization Tool (OVITO)[44].

The characteristic wavelength and amplitude of spinodal decomposition were quantified by analyzing concentration profiles along x-direction at multiple (y,z) locations within the simulation box. The wavelength was determined from the first nontrivial peak of the autocorrelation function of the concentration profile, which captures the dominant repeat distance of compositional fluctuations.[7,45,46] The amplitude was obtained by fitting the distribution of local concentrations to the Langer-Bar-on–Miller two-Gaussian mixture probability density function, thereby extracting the peak-to-trough composition contrast[47,48]. To account for statistical variation across different sampling locations and independent simulation runs, the median values of both wavelength and amplitude were reported, with associated uncertainties defined as half the interquartile range.

*Differential scanning calorimetry*

DSC measurements were performed on a Netzsch 404F3 Pegasus equipped with a silicon carbide furnace (maximum temperature 1650°C) and a type-P measurement head (maximum temperature 1150°C) for improved response at lower temperatures. DSC measurements were done under 99.999% argon cover gas with < 1 ppm oxygen impurities using a zirconium getter ring. For the validation test, two samples of Fe-36Cr aged at 500°C for 10 hours were measured. Samples were placed in manufacturer-provided alumina crucibles with lids. The temperature program consisted of five heating and cooling cycles, with a maximum temperature of 700°C (chosen to be above the approximate critical temperature of the spinodal phase transformation $T_{SP}$ ~ 627°C).[49] Dynamic steps were performed at a heating/cooling rate of 10°C/min, with an isothermal segment of 30 minutes at the minimum temperature of 100°C to allow thermal stabilization. DSC was performed with five heating/cooling cycles to 700°C at 10°C/min. The first heat captures the unmixing enthalpy, and subsequent heats serve as baseline.

The time-temperature history of DSC measurements is provided in the supplemental information file. To calibrate the instrument for the particular configuration of these experiments, corrections (e.g. measurements of the empty crucibles) were performed, as well as a measurement of a sapphire heat capacity standard in the temperature range of interest. A comprehensive description of calibration and corrections in DSC measurements is provided in the supplementary materials of ref. 50, which were closely followed here.

Because the enthalpy change of iron and chromium de-mixing is weak, post-processing steps are required, such as fitting the phase transformation peak as performed in refs. 51 and 52. First, the corrections are averaged and subtracted from the measurement of the combined sample and crucibles. Next, a third-order polynomial is fit to the DSC data outside of the region of interest for the phase transformation. A third-order polynomial fits the data well and is consistent with a heat capacity that depends only weakly on temperature beyond its second-order term. The region of interest for the phase transformation is 500°C - 625°C, based on the aging temperature, peak width, and heating rate combinations reported in other studies of unmixing.[51,52] The polynomial baseline is then subtracted from the DSC data. Next, since separation of phases formed during aging will only occur in the first heating step, the subsequent heats (two through five) are averaged across the two measured samples, to provide an “as-unmixed” value. Finally, the average of the first heating steps is taken, and the subsequent heating steps are subtracted, yielding a curve representing only the heat flow unique to the first heat.

*EBSD Rotation Matrix Calculations*

The Oxford Symmetry detector records electron backscatter diffraction patterns (EBSP) at each pixel in an EBSD map, along with other pertinent information. To find the surface orientation of the grain of interest, the Aztec software fits some number of bands on the EBSP and determines the relative relationship between the sample coordinate system and the crystal coordinate system, according to Bunge's convention. For each pixel, a set of three Euler angles (phi1, Phi, phi2) are recorded in degrees, and converted to radians for calculations. These angles define the relative rotations required to translate between the two coordinate systems. To compute the surface orientation in the crystal coordinate system from the known sample coordinate system, the rotation matrix needs to be computed for the set of Euler angles fit to the EBSP. The rotation scheme for Bunge's convention is a Z-X-Z' rotation for phi1, Phi, and phi2, respectively. In matrix notation, the three rotations can be defined as:

$$R_Z = \begin{bmatrix} \cos\phi_1 & \sin\phi_1 & 0 \\ -\sin\phi_1 & \cos\phi_1 & 0 \\ 0 & 0 & 1 \end{bmatrix} \tag{6}$$

$$R_X = \begin{bmatrix} 1 & 0 & 0 \\ 0 & \cos\Phi & \sin\Phi \\ 0 & -\sin\Phi & \cos\Phi \end{bmatrix} \tag{7}$$

$$R_{Z'} = \begin{bmatrix} \cos\phi_2 & \sin\phi_2 & 0 \\ -\sin\phi_2 & \cos\phi_2 & 0 \\ 0 & 0 & 1 \end{bmatrix} \tag{8}$$

The rotation matrix, $R$, is computed by the dot product of the three rotation matrices.

$$R = R_Z \cdot R_X \cdot R_{Z'} \tag{9}$$

The surface orientation is the last row of the 3x3 rotation matrix, $R$.

*Statistics and reproducibility*

For ex situ TGS, the standard deviation for each reported SAW speed at a given orientation is based on the fractional standard deviation of 10 measurements taken at one orientation within a single grain. Standard deviations in Table 1 were computed using SAW speeds measured over three orientations (spanning ≈12°) around the relative minimum described in Figure 3, which accounts for the ≈1-2° of uncertainty in locating the minimum across aging conditions. Elastic constant uncertainties in Table 2 are the 1-σ fitting errors from the inverse problem (ref. 21). Reproducibility of *ex situ* stiffening was confirmed across two independently prepared and characterized grains of Fe-36Cr (Figs. 3 and 6).

Vickers hardness values are reported as mean ± standard deviation over 9-11 indents per condition. Molecular dynamics/Monte Carlo elastic constants and densities are reported as mean ± standard deviation over 10 independent simulations with different initial random configurations, before and after thermal equilibration. Reported uncertainties denote standard deviations unless otherwise stated.

The statistical significance of the Fe-26Cr SAW softening was assessed with a two-tailed Welch's test (N = 3) for the difference of means ($p = 0.0164$), which does not assume equal variances, and a significance threshold of $\alpha = 0.05$ was used. No randomization or blinding was performed, as these are not applicable to the physical measurements and simulations reported here.

*Declaration of generative AI and AI-assisted technologies in the manuscript preparation process*

Generative AI and AI-assisted technologies (Claude Opus 4.8) were used to help proofread the manuscript for grammatical errors.

## Acknowledgments

The authors would like to convey sincere thanks to Prof. Linn Hobbs of MIT for providing the Fe-26Cr sample. We gratefully acknowledge Prof. Craig Carter for guidance and assistance in computing the characteristic spinodal wavelength.

## Funding

M.I. acknowledges support from the MathWorks Engineering Fellowship Fund. This work was performed in part at the Harvard University Center for Nanoscale Systems (CNS); a member of the National Nanotechnology Coordinated Infrastructure Network (NNCI), which is supported by the National Science Foundation under NSF award no. ECCS-2025158. Atomistic simulations were performed using the Expanse supercomputer at the San Diego Supercomputer Center through allocation MAT210005 from the Advanced Cyberinfrastructure Coordination Ecosystem: Services & Support (ACCESS) program, supported by National Science Foundation grants #2138259, #2138286, #2138307, #2137603, and #2138296. Additional computational resources were provided by the Extreme Science and Engineering Discovery Environment (XSEDE), supported by National Science Foundation grant #1548562. Throughout the duration of this work, S.A.A.D. was funded by the King Abdullah University of Science and Technology (KAUST) Fellowship and the MIT Department of Civil and Environmental Engineering (CEE) under the Friesecke (1961) Fellowship Fund. K.R. and T.G. acknowledge support from Czech Science Foundation [project No. 22-13462S] and from the project FerrMion [project No. CZ.02.01.01/00/22_008/0004591] of the Czech Ministry of Education, Youth, and Sports, co-funded by the European Union. The funders had no role in the study design, data collection and analysis, decision to publish, or preparation of the manuscript.

## Competing interests

The authors declare no competing financial or non-financial interests.

## Author Contributions

M.R. and M.P.S. conceived the study. M.S. contributed to the methodology for the atomistic and computational work. M.I. performed and analyzed the MD/MC simulations under the supervision of R.F. A.W. and A.J.A. performed the atom probe tomography, which M.R. and A.W. analyzed. A.N. performed and analyzed the differential scanning calorimetry, and wrote the supplemental information text. M.K.C. and A.K. performed the transient grating spectroscopy, which they analyzed together with M.R. T.G. wrote the code used for inverse elastic constant determination, and K.R. performed the corresponding computation and analysis. M.R. wrote the manuscript with assistance from M.K.C and M.I., and M.I., S.A.A.D., M.K.C., K.R., A.P.C.W., T.G., M.P.S., R.F., and A.J.A. reviewed and provided comments on the manuscript. All authors approved the final manuscript.

## Data Availability

The authors declare that the data supporting the findings of this study are available in the paper, the Supplementary Information file, and the GitHub repository for this paper (available in the public repository at https://github.com/shortlab/2025-mraechu-spinodal-is-real). The full-size APT file is not available due to file size, but will be made available upon request.

## Code Availability

The analysis code supporting this study is available in the public repository at https://github.com/shortlab/2025-mraechu-spinodal-is-real. The code used for inverse determination of the elastic constants from the surface acoustic wave dispersion is available in the public repository at https://github.com/tomasgrabec/GuliweRR.

## Figure Legends

Figure 1. Method of razor marking for successive TGS measurements and determination of grain surface orientation. Shows (a) Fe-36Cr after etching to reveal grain structure (optical image), (b) after razor marking and re-polishing (SEM image), and (c) a representative EBSD pattern. The red dashed outline in the optical image of the etched steel denotes the razor-outlined grain.

Figure 2. The effect of time at temperature on SAW speed. Shows (a) pictorial representation of time-temperature processing conditions for Fe-26Cr and Fe-36Cr, and (b) SAW speed at relative minimum for Fe-36Cr (error bars are one standard deviation above and below the mean). HOM indicates homogenization and HT stands for heat treatment.

Figure 3. SAW speed as a function of relative surface angle for Fe-36Cr and Fe-26Cr. Each point was taken in the same single grain after each heat treatment, and the standard deviation

of SAW speed measurements is ≈0.1% of the measured value, which is too small to see on the plot. Note that stiffening is directionally dependent. Slower directions typically exhibit more stiffening with thermal aging, while faster directions exhibit less, in both (a) the Fe-36Cr alloy, with purple, orange, grey, and blue markers for HOM1, HT1, HT2, and HOM2 conditions, respectively, and (b) the Fe-26Cr alloy with purple, grey, and blue markers for HOM1, HT1, and HOM2, respectively.

Figure 4. *In situ* TGS at ≈500°C shows clear progressive stiffening. The standard deviation of SAW speed measurements is ≈0.1% of the measured value, which is too small to be seen on the graph.

Figure 5. APT reconstruction results on Fe-36Cr aged for 10 hours at 500°C. Shows (a) a secondary electron image of the APT tip, (b) a 2D projection of a 38.6 at. % Cr 3D isosurface from a representative region of the tip, and (c) a proxigram for the same region.

Figure 6. TGS rotation study shows directionally dependent stiffening in the grain where APT was performed. Purple and orange markers are after the 1$^{st}$ homogenization and aging at 500°C for 10 hours, respectively.

Figure 7. Vickers hardness results as a function of aging time at 500°C for Fe-36Cr. Error bars denote one standard deviation in either direction.

Figure 8. $C'$ modulus as a function of chromium composition. Shown for EMTO (ref. 19) and empirical studies from literature (refs. 24–28) in blue and green markers, respectively, and from MD and TGS inverse extraction in this study, in red and purple markers, respectively.

Figure 9. OVITO visualizations of Fe-26Cr and Fe-36Cr in solid solution and thermally equilibrated (at 500°C) states used for elastic constant calculations. Red voxels denote iron atoms, and blue voxels denote chromium atoms, and the black arrows denote crystallographic direction.